%% file: roppp_final_1.tex
\documentclass[aip,jcp,preprint]{revtex4-2}
\usepackage{graphicx}
\usepackage{amssymb}
\usepackage{amsmath}
\usepackage{bm}
\usepackage{mathrsfs}
\usepackage{tcolorbox}
\usepackage{hyperref}
\usepackage{natbib}
\usepackage{bibentry}

\usepackage{fancyhdr}
\usepackage{mathrsfs}
\usepackage{relsize}
\usepackage{array}
\usepackage{booktabs}
\input{abbrev}

\newcommand{\bse}{\begin{subequations}}
\newcommand{\ese}{\end{subequations}}

\newcommand{\tabr}[1]{Table~\ref{tab:#1}}
\newcommand{\tabl}[1]{\label{tab:#1}}

\begin{document}
\bibliographystyle{tim}
\nobibliography*
\title{ExROPPP: Fast, Accurate and Spin-Pure Calculation of the Electronically Excited States of Organic Hydrocarbon Radicals}
\author{James D.\ Green}
\affiliation{Department of Chemistry, Christopher Ingold Building, University College London, WC1H 0AJ, UK}
\author{Timothy J.\ H.\ Hele}
\email{t.hele@ucl.ac.uk}
\affiliation{Department of Chemistry, Christopher Ingold Building, University College London, WC1H 0AJ, UK}
\date{\today}

\begin{abstract}
Recent years have seen an explosion of interest in organic radicals due to their promise for highly efficient organic light-emitting diodes (OLEDs) and molecular qubits. However, accurately and inexpensively computing their electronic structure has been challenging, especially for excited states, due to the spin-contamination problem. Furthermore, while alternacy or ‘pseudoparity’ rules have guided the interpretation and prediction of the excited states of closed-shell hydrocarbons since the 1950s, similarly general rules for hydrocarbon radicals have not to our knowledge yet been found. In this article we present solutions to both of these challenges. We firstly combine the extended configuration interaction singles (XCIS) method with Pariser-Parr-Pople (PPP) theory to obtain a method which we call ExROPPP (Extended Restricted Open-shell PPP theory). We find that ExROPPP computes spin-pure excited states of hydrocarbon radicals with comparable accuracy to experiment as high-level GMC-QDPT calculations but at a computational cost that is at least two orders of magnitude lower. We then use ExROPPP to derive widely-applicable rules for the spectra of alternant hydrocarbon radicals which are completely consistent with our computed results. These findings pave the way for the highly accurate and efficient computation and prediction of the excited states of organic radicals.
\end{abstract}

\maketitle

\section{Introduction}
\label{sec:intro}
\pagenumbering{arabic}
It was long thought that the search for stable emissive organic radicals was hopeless, as none were known experimentally.\cite{hir87a} However, since the discovery of stable, highly emissive organic radicals and the development of highly emissive radical OLEDs, there has been an explosion of interest in the field.\cite{pen15a,ai18a,guo19a,zhi20a,li22a,mur23a,cho23a} External quantum efficiencies in excess of 25\% and internal quantum efficiencies of near unity\cite{ai18a} for some radical-based devices have been reported. Although phosphorescent OLEDs based on molecules containing heavy elements and those based on organic closed-shell molecules which exhibit thermally activated delayed fluorescence also have high efficiencies, such figures are usually hard to achieve in conventional closed-shell OLEDs due to spin-statistics.\cite{bal98a,bal99a,fre99a,wan12a,min14a,shi08a,jan14a} Such high-performing OLEDs have potential applications in the next generation of lightweight and flexible displays and lighting.\cite{ai18a} Moreover, radical-based devices can achieve this intense emission in the near infrared (NIR) region, a characteristic which is unusual for small molecules and highly desirable for the development of fluorescent probes and NIR sources for medical diagnostic applications.\cite{guo14a,bla02a,sak16a,kon08a} Emissive organic radicals have since been incorporated into emissive conjugated polymers,\cite{mur23a} which opens up the possibility of having the efficiency and spectral benefits of a radical emitter with the processability advantages of conjugated polymers and also the possibility of semiconductivity.\cite{fre99a,cla10a} In addition, emissive organic radicals have potential applications as spin-optical interfaces in quantum information processing.\cite{gor23a} 

While there has been considerable work over the years to investigate the electronic structure of radicals, diradicals and polyradicals,\cite{stu19a,lon50a,lon55a} accurately computing their electronic structure remains challenging, largely due to spin-contamination. The first problem is whether to use unrestricted or restricted orbitals,
as the most obvious choice which gives the most accurate energies and the best description of the ground state are unrestricted orbitals, but using them results in spin-contamination. The second problem is the choice of determinants for the excited states, as including all single-excitations in configuration interaction singles (CIS) results in spin-contaminated excited states, even when using a restricted basis.\cite{lon55a,mau96a,roe13a,hel21a}
Furthermore, while general alternacy or pseudoparity rules have been known since the 1950s\cite{par56a} to assign, predict and interpret the spectra of ground-state closed-shell hydrocarbons, to the best of our knowledge no such general rules exist for the excited states of hydrocarbon radicals.

In this article, focusing on organic radicals with a single unpaired electron (monoradicals), we give a detailed overview of some of the existing theories for radical electronic structure calculation, the challenges that arise when it comes to excited states of radicals and how or if they can be overcome. We then expand on this previous work seeking to provide new insight into the properties of the excited electronic states of radicals.
 
One problem encountered in the calculation of the electronically excited states of radicals is that there is no ideal set of orbitals to choose from. Both the unrestricted Hartree-Fock (UHF) method\cite{pop54a} and the restricted open-shell Hartree-Fock (ROHF) method\cite{roo60a,huz61a,dav73a,bin74a} have their shortfalls. While UHF is often preferred as it gives lower, more accurate energies due to each of the electrons having different spatial orbitals, UHF states comprising a finite linear combination of UHF determinants are generally not eigenfunctions of the total spin operator $\hat S^2$.\cite{sza89a} Though UHF (and Unrestricted Density Functional Theory) is a useful tool for calculating accurate ground state properties and in many such cases the spin-contamination can be ignored\cite{bar93a,mal00a,ras09a,cab97a,nee06a,mue81a,stu19a,roo16a,bak93a}, it can be problematic for excited states where spin contamination is unavoidable and can lead to unphysical or uninterpretable results.\cite{lon55a,hel21a,man05a,ipa09a,li11a,stu19a}. Conversely, ROHF eliminates spin-contamination by construction for states where all unpaired electrons are of the same spin, including the ground state. Furthermore, it is possible to form spin-adapted ROHF configurations in a finite number of determinants for all other states.\cite{lon55a,roo60a,huz61a,dav73a,bin74a,sza89a,roo16a,hel21a} However, ROHF has its weaknesses, namely that it gives less accurate energies than UHF and also the issue of orbital energies not being uniquely defined.\cite{gla10a,tak10a}

Although both UHF and ROHF clearly have their own limitations, for the calculation of the photophysical properties of radicals we believe that spin-purity must take precedence over a slight improvement in accuracy. However, using a restricted reference is not on its own a sufficient criterion for satisfying spin-purity of the excited states, as one needs to ensure that the correct linear combinations of determinants are chosen for states with more than one open-shell.

This brings us to the second major challenge of preserving spin-purity of the excited states. In the case of closed-shell molecules, from the set of all single-excitations of $\alpha$ and $\beta$ electrons, $\ket{\Phi_i^{j^\prime}}$ and $\ket{\Phi_{\bar{i}}^{\bar{j}^\prime}}$ respectively, one can form a complete set of spin-adapted singlet $\ket{^1\Psi_i^{j^\prime}}$ and triplet $\ket{^3\Psi_i^{j^\prime}}$ configurations or configuration state functions (CSFs) from a restricted determinant in configuration interaction singles (CIS). This is the simplest configuration interaction method and the formal starting point for many other more accurate methods.\cite{sza89a,roo16a,kim22a} Usually we are only interested in the singlet states with regard to absorption and emission as triplet states are formally dark. Caution must be taken, however, when choosing a basis for the excited states of radicals. Firstly, if one were to use a basis of all single excitations, as in the closed-shell case, and form a complete set of states, this would in general result in spin-contamination. This is demonstrated pictorially in \figr{spin} below, and also numerically \tabr{benzyl_comp} where we have presented the results of this approach for the benzyl radical. The general spin properties of closed-shell molecules and radicals are summarized in \tabr{csos}.

One solution to the spin-contamination problem in radical CIS is to restrict the basis of single excitations to only include those which are pure doublets, which are those which have only one open-shell, in which case the resulting states would also be pure doublets. Alternatively, one could go a step further and also add in some spin-adapted doublet states for those configurations containing three open-shells,\cite{mau95a} however, neither method captures quartet states. Furthermore, some excited doublet states with three open-shells are also not captured by these methods, which, despite being dark, could affect  the spectrum. This problem is solved if one also includes a certain type of double excitations in the basis, as shown in previous literature, where in this case a complete set of spin-pure doublet and quartet states can be constructed on the same theoretical footing, in a similar manner to closed-shell CIS.\cite{lon55a,mau96a} This formulation of open-shell CI, which is discussed further in Section~\ref{sec:bkgnd} is sometimes referred to as extended CIS or XCIS\cite{mau96a}. This is different to CISD where \emph{all} double excitations are included and which would not in general be spin-pure for radicals. Also XCIS has the same scaling as CIS, where the number of determinants scales with $\mathcal{O}(N_{\mathrm{occ}}N_{\mathrm{vir}})$ whereas CISD scales with $\mathcal{O}(N_{\mathrm{occ}}^2 N_{\mathrm{vir}}^2)$, where $N_{\mathrm{occ}}$ is the number of occupied orbitals and $N_{\mathrm{vir}}$ the number of virtual (unoccupied) orbitals in the ground state.
\begin{figure}
    \centering
    \includegraphics[width=0.5\textwidth]{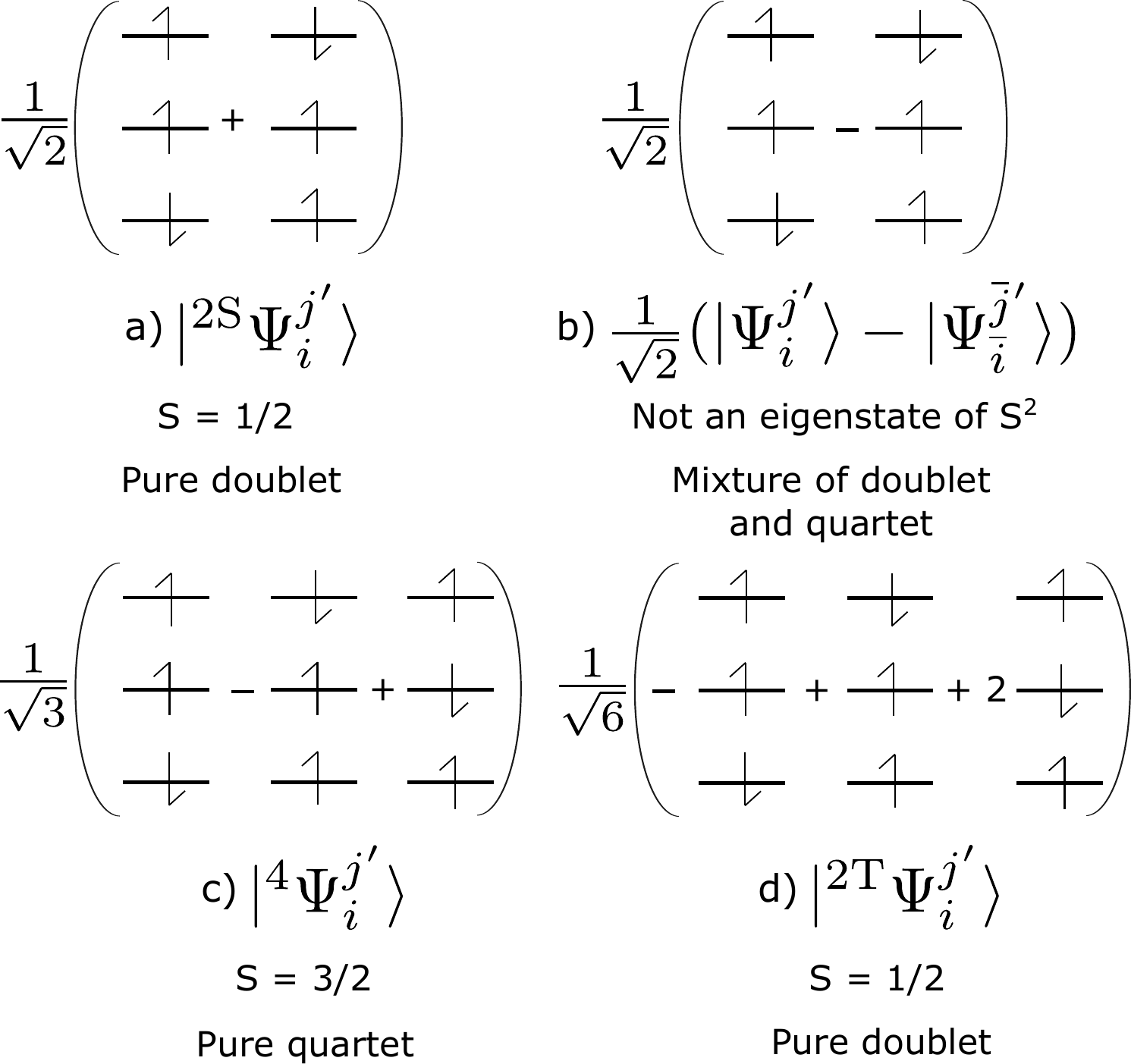}
    \caption{Illustration of the problem of using only single excitations to describe the excited states of a doublet radical which consist of three open-shells. Four excited configurations are pictured: a) the spin-pure singlet-coupled doublet (2S) state (`+' combination) formed of only single excitations, b) the `$-$' combination of two single excitations which is spin-contaminated, c) the spin-pure quartet state composed of single and double excitations and d) the spin-pure triplet-coupled doublet (2T) state composed of of single and double excitations.}   
    \figl{spin}
\end{figure}

\begin{table}[h]
\caption{Comparison of the general spin properties of the singly-excited electronic states for closed-shell molecules and radicals.}
\tabl{csos}
\begin{tabular}{c|c|c}
Property& Closed-shell & Radical \\ \hline
Ground state multiplicity & Singlet & Doublet \\
Determinant & RHF & ROHF or UHF \\
Spin-multiplicity of single-excitations &$\ket{\Psi_i^{j^\prime}} + \ket{\Psi_{\bar i}^{\bar j^\prime}}$ is singlet. & $\ket{\Psi_{\bar i}^{\bar 0}}$, $\ket{\Psi_0^{j^\prime}}$, $\ket{\Psi_i^{j^\prime}} + \ket{\Psi_{\bar i}^{\bar j^\prime}}$\\
 & $\ket{\Psi_i^{j^\prime}} - \ket{\Psi_{\bar i}^{\bar j^\prime}}$ is triplet. & are doublets. \\
 && $\ket{\Psi_i^{j^\prime}} - \ket{\Psi_{\bar i}^{\bar j^\prime}}$ contaminated. \\
Usual multiplicity of  & Triplet& Doublet\\
lowest excited state& & \\
Spin restriction alone sufficient & Yes & No \\
for spin-purity?& & 
\end{tabular}
\end{table}

We must stress here that there are many other existing methods to obtain accurate, spin-pure excited states of radicals. There are the multiconfigurational SCF (MCSCF) method, multiconfigurational second-order perturbation theory such as CASPT2 and GMC-QDPT mentioned later, full-CI (FCI), coupled-cluster (CC) theory and so on.\cite{wer81a,mal89a,nak93a,sza89a,mar23a} These methods achieve spin-pure results for radicals by using a restricted open-shell reference and by either including all possible configurations (within a given active space in MCSCF and CASPT2/GMCQDPT) or including fewer than all possible configurations and employing spin-adaption techniques.\cite{iva03a,bro08a,sza00a} When used correctly these methods can give accurate results,\cite{sch08a} although for very accurate excitation energies one must also include nuclear quantum effects.\cite{hel21b} However, these calculations are computationally expensive, their computational cost increasing by many orders of magnitude with increasing molecular size, and also are by no means black-box methods, requiring chemical intuition and experience to accurately utilize.\cite{sza89a} One potential avenue is to decrease the computational cost of such methods by, for example, combining MCSCF or FCI with quantum Monte Carlo, however, discussions on this topic are beyond the scope of this article.\cite{boo10a,tho15a} On the other hand, time-dependent density functional theory (TD-DFT) allows for the rapid calculation of electronic excited state spectra which does not require any non-trivial input from the user and has become a popular method for the simulation of UV/Visible spectra.\cite{sil08a,lau13a} However, TD-DFT has several drawbacks, namely that: (a) its results can vary significantly depending on the chosen functional \cite{sil08a,lau13a}, (b) it is inaccurate at describing long-range phenomena such as charge-transfer without the use of specifically tailored long-range functionals,\cite{dre03a,pea08a,ste09a,ali20a} and (c) it does not typically give spin-pure states, as an unrestricted version of the Kohn-Sham equations is often preferred in favour of more accurate energies, and even if a restricted open-shell reference is used, conventional TD-DFT does not form spin-adapted excited states.\cite{ipa09a,li11a} A spin-adapted version of TD-DFT for radicals has recently been developed, called X-TDDFT, which is clearly one faster alternative to the more expensive MCSCF/CASPT2/GMC-QDPT/FCI/CC methods.\cite{li11a} However, X-TDDFT is by no means theoretically straightforward since the required double excitations are not compatible with conventional TD-DFT so a work-around using spin-flip single excitations and reference states with different values of $M_s$ needs to be implemented. Finally, although TD-DFT is faster than the post-Hartree-Fock methods mentioned above, even it becomes unfeasibly expensive for large numbers of calculations, and there is a growing need for ever faster property determination methods for high-throughput virtual screening of molecular candidates and AI algorithm search methods, and for property determination of large molecules. \cite{gom16a,yu23a,gre22a}

This brings us to two key aims which we will address in this paper: 
 \paragraph{An inexpensive method for screening radicals:}
 What is the simplest and least computationally expensive method which can calculate spin-pure doublet and quartet states of radicals with reasonable accuracy?
 \paragraph{Design rules for emissive radicals based on alternant hydrocarbons:} 
 A general set of rules for the interaction of the excited states of radicals would allow for a better understanding of their UV/Visible spectra, guiding the assignment of bands and allowing us to predict the effect of structural variations on their spectra. Considering that many emissive radicals are (non-alternant) derivatives of alternant hydrocarbons such as the triphenylmethyl radical,\cite{pen15a,ai18a,guo19a,zhi20a,li22a,mur23a} a good starting point would be to deduce such design rules for alternant hydrocarbon radicals.
 
The article is structured as follows. Firstly, in Section~\ref{sec:bkgnd} we define the key theoretical concepts which will inform the discussion of our methodology. Here we discuss the XCIS method for doublet radicals,\cite{lon55a,mau96a} a restricted open-shell method for the molecular orbitals,\cite{lon55a} Pariser-Parr-Pople theory and its approximations\cite{par53a,pop53a,pop55a,par56a, kou67a} and also give a brief overview of the theories relating to alternant hydrocarbons.\cite{cou40a,pop55a,par56a} In Section~\ref{sec:mthd} we show how XCIS and PPP theory with a restricted open-shell determinant can be combined and implemented in a method we call ExROPPP, for the rapid calculation of spin-pure exicted states of radicals which is not limited to just alternant hydrocarbons. The numerical results of computations with ExROPPP are presented in Section~\ref{sec:exroppp_results} where the method is benchmarked against the highly accurate multiconfigurational perturbation theory method GMC-QDPT and experimental data, where ExROPPP is seen to have a similar accuracy to GMC-QDPT. Finally, in Section~\ref{sec:alt_rules} we then apply the algebra of XCIS to alternant hydrocarbons and by making a small number of approximations, we derive rules for the interaction of excited states of alternant hydrocarbon radicals similar to those derived previously for closed-shell molecules\cite{par56a}. We believe that these rules may aid with the assigment of bands in the UV/Visible spetcra of radicals and also guide the design of emissive radicals. 

\section{Background Theory}
\label{sec:bkgnd}
Here we discuss the XCIS method for the excited states of radicals,\cite{lon55a,mau96a} the restricted open-shell method of Longuet-Higgins and Pople\cite{lon55a} which allows us to obtain self-consistent molecular orbitals for a system with one unpaired electron in its ground state, the various approximations of Pariser-Parr-Pople (PPP) theory,\cite{par53a,pop53a,pop55a,par56a, kou67a} and finally the theories of Coulson, Rushbrooke, Pople and Pariser which apply alternant hydrocarbons which we will call upon to derive alternacy rules for radicals.\cite{cou40a,pop55a,par56a}

\subsection{From CIS to XCIS}
\label{sec:xcis}
We start by discussing the description of the ground state of the radical. Restricted orbitals are chosen where $k$ doubly-occupied closed-shells and one singly-occupied open-shell make up the ground state, which is a pure doublet with $S = M_S = 1/2$. We label the orbitals choosing $0$ to be the singly-occupied molecular orbital (SOMO), and successively number the doubly occupied orbitals in order of descending energy starting from $1$ (the HOMO), and the vacant orbitals in order of ascending energy with primes starting from $1^\prime$ (the LUMO). 
We choose this notation as it is the same as that used by Pariser\cite{par56a} and Ref.~\citenum{hel21a}. We now discuss the excited configurations. Single excitations from the ground state $\ket{^2\Psi_0}$ give rise to the configurations: $\ket{^2\Psi_{\bar i}^{\bar0}}$, $\ket{^2\Psi_0^{j^\prime}}$, $\ket{\Psi_i^{j^\prime}}$ and $\ket{\Psi_{\bar i}^{\bar j ^\prime}}$ (see Figure I in the supplementary information for a comprehensive molecular orbital diagram).

As we are concerned with finding those states which are eigenstates of the total spin operator $\hat S^2$, we shall ascertain the effect of $\hat S^2$ on the above configurations. States $\ket{^2\Psi_0}$,  $\ket{^2\Psi_{\bar i}^{\bar0}}$ and $\ket{^2\Psi_0^{j^\prime}}$ are eigenstates of $\hat S^2$ where $S=1/2$ --- these configurations are pure doublet states. This is to be expected as they have only one open-shell.
On the other hand, those configurations which have three open-shells, $\ket{\Psi_i^{j^\prime}}$ and $\ket{\Psi_{\bar i}^{\bar j ^\prime}}$, are not eigenstates of $\hat S^2$ and are mixtures of doublet and quartet states. Naturally we can make two orthonormal linear combinations of $\ket{\Psi_i^{j^\prime}}$ and $\ket{\Psi_{\bar i}^{\bar j ^\prime}}$, 
\bse
\begin{align}
\ket{^{2\mathrm{S}}\Psi_i^{j^\prime}} &= \frac{1}{\sqrt 2}(\ket{\Psi_i^{j^\prime}} + \ket{\Psi_{\bar i}^{\bar j ^\prime}}) \eql{ijp}\\
\ket{^{\mathrm{DQ}}\Psi_i^{j^\prime}} &= \frac{1}{\sqrt 2}(\ket{\Psi_i^{j^\prime}} - \ket{\Psi_{\bar i}^{\bar j ^\prime}}) \eql{ijm}
\end{align}
\ese
to see if we can construct two spin-adapted states. We find that the `$+$' combination is a pure doublet with $S=1/2$, however, the `$-$' combination is not an eigenstate of $\hat S^2$. It is clearly impossible then to construct a spin-pure basis for doublet and quartet excited states of a radical from just single excitations from the ground state.

As previous literature has shown, one must therefore go to higher-order excitations to attempt to satisfy spin-purity whilst calculating doublet and quartet states on the same theoretical footing.\cite{lon55a,mau96a} Double excitations of the type $\ket{\Psi_{\bar{i}0}^{\bar0j^\prime}}$ were suggested as, along with the single-excitations $\ket{\Psi_i^{j^\prime}}$ and $\ket{\Psi_{\bar i}^{\bar j ^\prime}}$, they complete all permutations of three open-shells in orbitals $i$, $0$ and $j^\prime$ with $M_S=1/2$.

One can construct the matrix representation of $\hat S^2$ in the basis of three excitations: $\ket{\Psi_i^{j^\prime}}$, $\ket{\Psi_{\bar i}^{\bar j ^\prime}}$ and $\ket{\Psi_{\bar i 0}^{\bar 0 j ^\prime}}$
\begin{align}
\mathrm{\mathbf{S^2}} = 
\begin{pmatrix}
7/4 & -1 & 1\\
-1 & 7/4 & -1\\
1 & -1 & 7/4 \\
\end{pmatrix}
\end{align}
which has eigenvalues of 3/4 twice and 15/4, corresponding to two doublet eigenstates and one quartet eigenstate 
\bse
\begin{align}
\ket{^{2\mathrm{S}}\Psi_i^{j^\prime}} &= \frac{1}{\sqrt2}(\ket{\Psi_i^{j^\prime}} + \ket{\Psi_{\bar i}^{\bar j^\prime}}) \eql{2s_def}\\
\ket{^{2\mathrm{T}}\Psi_i^{j^\prime}} &= \frac{1}{\sqrt6}(-\ket{\Psi_i^{j^\prime}} + \ket{\Psi_{\bar i}^{\bar j^\prime}}+2\ket{\Psi_{\bar i 0}^{\bar0j^\prime}}) \eql{2t_def} \\
\ket{^4\Psi_i^{j^\prime}} &= \frac{1}{\sqrt3}(\ket{\Psi_i^{j^\prime}} - \ket{\Psi_{\bar i}^{\bar j^\prime}}+\ket{\Psi_{\bar i 0}^{\bar0j^\prime}}). \eql{quart_def}
\end{align}
\ese
These three states then, along with $\ket{^2\Phi_0}$, $\ket{^2\Psi_{\bar i}^{\bar0}}$ and $\ket{^2\Psi_0^{j^\prime}}$, constitute a basis for the spin-pure excited doublet and quartet states of a radical and will be the starting point for our methodology.\cite{lon55a,mau96a,hel21a} We note here that while the quartet eigenstate $\ket{^4\Psi_i^{j^\prime}}$ is uniquely defined, the two doublet eigenstates are not, since they are spin degenerate and can be mixed. 
The first doublet state [Eq.~(\ref{eq:2s_def})] we label 2S as it can have a nonvanishing dipole moment from the ground state\cite{lon55a,hel21a} and is sometimes referred to as `singlet-coupled'\cite{he19a} and the second doublet state [Eq.~(\ref{eq:2t_def})] we refer to as 2T as it is always completely dark and is sometimes referred to as `triplet-coupled'\cite{he19a}.

\subsection{The restricted open-shell method for the orbitals}
\label{sec:rohf}
 While there are many variations of restricted open-shell methods to choose from, we elect to use the method of Longuet-Higgins and Pople.\cite{lon55a} In this method, the ground state orbitals are solved for variationally as per usual HF, but in this case minimising the ground state energy with respect to mixing of small amounts of single excitations $\ket{\Psi_{\bar i}^{\bar 0}}$, $\ket{^2\Psi_0^{j^\prime}}$ and $\ket{^{2\mathrm{S}}\Psi_i^{j^\prime}}$ into the ground state. 


They propose a single Fock operator for both closed and open-shell spinorbitals
\begin{align}
\bra{\psi_p}\hat{H}\ket{\psi_q}=F_{pq} &= h_{pq} + \sum_{l=1}^{k}[2(pq|ll)-(pl|lq)] + \frac{1}{2}[2(pq|00)-(p0|0q)] \eql{fop}.
\end{align}
In this method, the ground state mixing of the configuration $\ket{^{2\mathrm{S}}\Psi_i^ {j^\prime}}$ is exactly zero however, the ground state mixing of the configurations $\ket{^2\Psi_{\bar i}^{\bar 0}}$ and $\ket{^2\Psi_0^{j^\prime}}$ is \emph{almost} but not exactly zero (see supplementary information for further details). 
Therefore, Brillouin's theorem is \emph{almost} but not completely satisfied. As a note here, there are other restricted open-shell approaches for the orbitals which are more general in that they can be applied to systems with more than one open-shell in the ground state, and which also satisfy Brillouin's theorem exactly.\cite{roo60a,huz61a,dav73a,bin74a} However, these more general methods involve partitioning the set of orbitals into interacting subspaces for doubly-occupied, partially-occupied and vacant orbitals each with their respective Fock operators, requiring the use of Lagrange multipliers to solve the coupled Roothaan equations whilst maintaining spin-restriction.\cite{roo60a,huz61a,dav73a,bin74a} Whilst this ensures Brillouin's theorem is satisfied exactly, these methods are complicated to implement and we therefore elect to use the approximate method of Longuet-Higgins and Pople for systems with a single open-shell doublet ground state, despite it only approximately satisfying Brillouin's theorem, as it is theoretically simpler and only involves diagonalizing a single Fock matrix for all orbitals.  As we shall see later, this simple form of the Fock operator also allows the derivation of elegant rules for the spectra of alternant hydrocarbon radicals. 



\subsection{Pariser-Parr-Pople theory}
\label{sec:ppp}
Thus far, in our discussion of the method to obtain the molecular orbitals, we have assumed an arbitrary basis of atomic orbitals and have not detailed the method to be used to obtain the one-electron (core) integrals $h_{\mu \nu}$ and two-electron integrals $(\mu \nu| \rho \sigma)$ needed to form the Fock matrix. We could elect to use Gaussian orbitals and compute the integrals for the set of all atomic orbitals exactly, however, this comes at a computational expense which we aim to minimise. 
We therefore call upon Pariser-Parr-Pople (PPP) theory\cite{par53a,pop53a,pop55a,par56a, kou67a} as it is arguably the simplest and least computationally expensive method which is known to accurately reproduce the excited state spectra of $\pi$-conjugated closed-shell molecules,\cite{hel19a,alv19a} and we will see that it performs similarly well for $\pi$-conjugated organic radicals when using restricted orbitals and coupled with XCIS in a method we call ExROPPP. 
Moreover, due to the approximations within PPP theory, many integrals vanish and many different integrals become equivalent, simplifying the form of the Hamiltonian and other electron operators. Therefore, applying the approximations of Pariser-Parr-Pople (PPP) theory to the Hamiltonian and dipole moment operators within XCIS allows us to derive alternacy rules. 

PPP theory is a semiempirical method for the electronic structure calculation of conjugated $\pi$ systems. The $\pi$ system is treated using an implicit basis of $p$ orbitals centred on the atomic nuclei, where neighbouring $p$ orbitals interact via a H{\"u}ckel $\beta$ (resonance integral) term. The $\sigma$ system is treated as core-like and only interacts with the $\pi$ system through a H{\"u}ckel $\alpha$ (core integral) term which also accounts for nuclear charge.

In PPP theory, the neglect of differential overlap (NDO) approximation  is applied, where the overlap integral between any two atomic orbitals is equal to the Kronecker delta function of the two orbitals: $\langle \phi_\mu| \phi_\nu \rangle= \delta_{\mu \nu}$.
This firstly means that the Roothaan equations simplify to $\mathrm{\mathbf{FC}} = \mathrm{\mathbf{CE}}$,
where $\mathbf{E}$ is a diagonal matrix of orbital energies $\{\epsilon_p\}$. 
The second implication of NDO is that the expressions for the two-electron integrals are greatly simplified, as those integrals which depend on the overlap of two $p$ orbitals on different atoms are neglected. Therefore, the two-electron integrals take the form:
\begin{align}
(\mu\nu|\rho\sigma) = (\mu\mu|\rho\rho)\delta_{\mu \nu}\delta_{\rho \sigma} =\gamma_{\mu \rho}\delta_{\mu \nu}\delta_{\rho \sigma},
\end{align}
where $\gamma_{\mu \rho}$ is a function of the distance between the two atomic centres.\cite{hel21a} 

\subsection{Alternant hydrocarbon radicals}
\label{sec:alt}
 Alternant hydrocarbons are conjugated hydrocarbons whose atoms can be divided into two sets, often referred to as \emph{starred} and \emph{unstarred}, where no two atoms of the same set may be directly bonded. This structural symmetry results in underlying symmetries in the energies and coefficients of pairs of orbitals, and degeneracy of pairs of single-excitations. This pairing theorem was first proven by Coulson and Rushbrooke for closed-shell systems, where it was shown to hold under the approximations of neglect of differential overlap (NDO) and resonance interaction only between nearest neighbouring atoms, such approximations being present in Pariser-Parr-Pople theory (see Section~\ref{sec:ppp}).\cite{cou40a} Subsequently, the Coulson-Rushbrooke pairing theorem was also shown to hold for odd alternant hydrocarbon radicals.\cite{lon55a,dew68a} 
 This symmetry of alternant hydrocarbons are sometimes referred to as pseudoparity, particle-hole symmetry or the Coulson-Rushbrooke-Longuet-Higgins theorem.\cite{cal83a,ziv88a,kar89a} 
\subsubsection{Orbital energy symmetry}
According to the theorem, for every bonding orbital there exists an antibonding orbital with an equal and opposite energy relative to the energy of a non-interacting atomic orbital. In the case of monoradicals, the SOMO has non-bonding character and its energy is the same as that of a non-interacting orbital, and all \emph{other} orbitals can be grouped into bonding and antibonding pairs which are equally spaced in energy about the SOMO: $(\epsilon_i + \epsilon_{i^\prime})/2=(F_{i,i} + F_{i^\prime, i^\prime})/2=\epsilon_0 = F_{0,0}.$
\subsubsection{Orbital coefficient symmetry}
 In alternant hydrocarbon radicals the SOMO has amplitude on only the \emph{starred} atoms, and for every Coulson-Rushbrooke orbital pair, the orbital coefficients in bonding and antibonding pairs: i) are equal in magnitude, and ii) have equal signs on the \emph{starred} atoms and opposite signs on the \emph{un-starred} atoms. In the convention followed in this paper, for odd alternant hydrocarbon radicals the number of \emph{starred} atoms is one greater than the number of \emph{unstarred} atoms. 
 The following is then true for two electron integrals for alternant hydrocarbon radicals: $(ij|kl) = (ij|k^\prime l^\prime) = (i^\prime j^\prime|kl) = (i^\prime j^\prime|k^\prime l^\prime)$, $(i^\prime j|kl) = (ij^\prime|kl)$ and $(i0|jk) = (i^\prime0|jk)$.
The latter relation is a special case of the first two because the SOMO is unchanged by changing the signs of the \emph{un-starred} atoms, i.e. $\psi_0=\psi_{0^\prime}$. Similar identities also exist for dipole moments: $\mu_{ij} = \mu_{i^\prime j^\prime}$, $\mu_{ij^\prime} = \mu_{i^\prime j}$ and $\mu_{i0} = \mu_{i^\prime 0}$, where we use the standard definition of the dipole moment in atomic units $\mu_{ij} = \bra i\hat \mu\ket j = -\bra i\bmr \ket j$.

\subsubsection{Degeneracy of the excited states with a single open-shell} These symmetry relations result in degeneracies in pairs of excitations, even when two electron terms are taken into account within the regime of the above approximations. The excitations $\ket{^2\Psi_{\bar i}^{\bar0}}$ and $\ket{^2\Psi_0^{i^\prime}}$ are degenerate for corresponding pairs of bonding and antibonding orbitals $\psi_i$ and $\psi_{i^\prime}$. They can be factored into two-determinant linear combinations with a `$+$' or `$-$' sign, shown below in \eqr{ps0}.\cite{lon55a,hel21a}
\begin{align}
\ket{^2\Psi_{0i}^\pm} = \frac{1}{\sqrt 2}(\ket{^2\Psi_{\bar{i}}^{\bar{0}}} \pm \ket{^2\Psi_0^{i^\prime}}). \eql{ps0} 
\end{align}

\section{Methodology}
\label{sec:mthd}

We introduce a novel method, which we call ExROPPP (\textbf{Ex}tended \textbf{R}estricted \textbf{O}pen-shell \textbf{P}ariser-\textbf{P}arr-\textbf{P}ople theory) for the rapid and accurate simulation of spin-pure excited states of organic $\pi$-conjugated radicals. This method marries the idea of XCIS\cite{lon55a,mau96a} with PPP theory\cite{par53a,pop53a,pop55a,par56a, kou67a} using the restricted open-shell method of Ref.~\citenum{lon55a}. We implement the equations in Sections~\ref{sec:rohf} and \ref{sec:ppp} in an in-house code, using the same parameters as those used previously for closed-shell molecules,\cite{hel19a,gre22a} to form the restricted open-shell Fock matrix in PPP theory and to subsequently obtain the optimized molecular orbital coefficients and orbital energies by solving the Roothaan equations iteratively until convergence of the total energy. The essential equations of XCIS discussed in Section~\ref{sec:xcis} (see $\mathrm{\mathbf{H_{\textrm{XCIS}}^x}}$, $\mathrm{\mathbf{H_{\textrm{XCIS}}^s}}$, $\mathrm{\mathbf{M_{\textrm{XCIS}}^x}}$ and $\mathrm{\mathbf{M_{\textrm{XCIS}}^s}}$ in SI Tables IX, X, XIII and XIV) are then implemented to form and diagonalize the XCIS Hamiltonian to give the energies and expansion coefficients of the XCIS excited states, and to compute their dipole moments and oscillator strengths for the plotting of linear absorption spectra. This method was benchmarked on a series of alternant hydrocarbon radicals and the results of these computations can be found in Section~\ref{sec:result}. However, we must stress that this method is not limited to just alternant hydrocarbon radicals but can be applied to any radical with one unpaired electron in its ground state. 

\section{Results}
\label{sec:result}
In this section we now present the central results of this paper. First we present the results of numerical calculations on a range of alternant hydrocarbon radicals with our new method: ExROPPP, where ExROPPP is shown to be accurate in comparison to experimental data and higher-level methods, spin-pure and computationally inexpensive. Secondly, we reveal simple and elegant rules for the interaction of the excited states of alternant hydrocarbon radicals, similar to those found by Pariser for closed-shell molecules\cite{par56a} which we believe could aid the design of radicals.

\subsection{Numerical results from ExROPPP on alternant hydrocarbon radicals}
\label{sec:exroppp_results}
Here we present the results of our new method: ExROPPP, benchmarked against experimental data and the results of calculations with General Multi-Configurational Quasi-Degenerate Perturbation Theory (GMC-QDPT), an accurate post-Hatree-Fock method for calculating excited states. The UV/Visible vertical absorption spectra of a series of odd alternant hydrocarbon radicals pictured in \figr{mols} were simulated using the ExROPPP method described in Section~\ref{sec:mthd}, and these results were also compared to the results of calculations using GMC-QDPT and experimental spectroscopic data in the literature where available. We must stress here that we choose to use alternant hydrocarbons to demonstrate the utility of ExROPPP, despite the fact ExROPPP is not limited to just alternant hydrocarbons, as they are generally stable and thus there is a large amount of experimental spectroscopic data available for them in the literature. Also, since many emissive radicals are non-alternant derivatives of alternant hydrocarbons, the results of these computations are still of relevance to the design of emissive radicals. The ExROPPP and GMC-QDPT simulated spectra of the radicals are presented together in figures \ref{fig:allyl} -- \ref{fig:tripxen} and their spectroscopically observed states computed by the two theoretical methods compared to the experimental data are presented in Tables~\ref{tab:allyl} -- \ref{tab:txm}, with further detail in Tables I -- VII of the SI. The results of the two theoretical methods are generally consistent with each other and closely reproduce the experimental excitation energies. 
\begin{figure}
\includegraphics[width=0.5\textwidth]{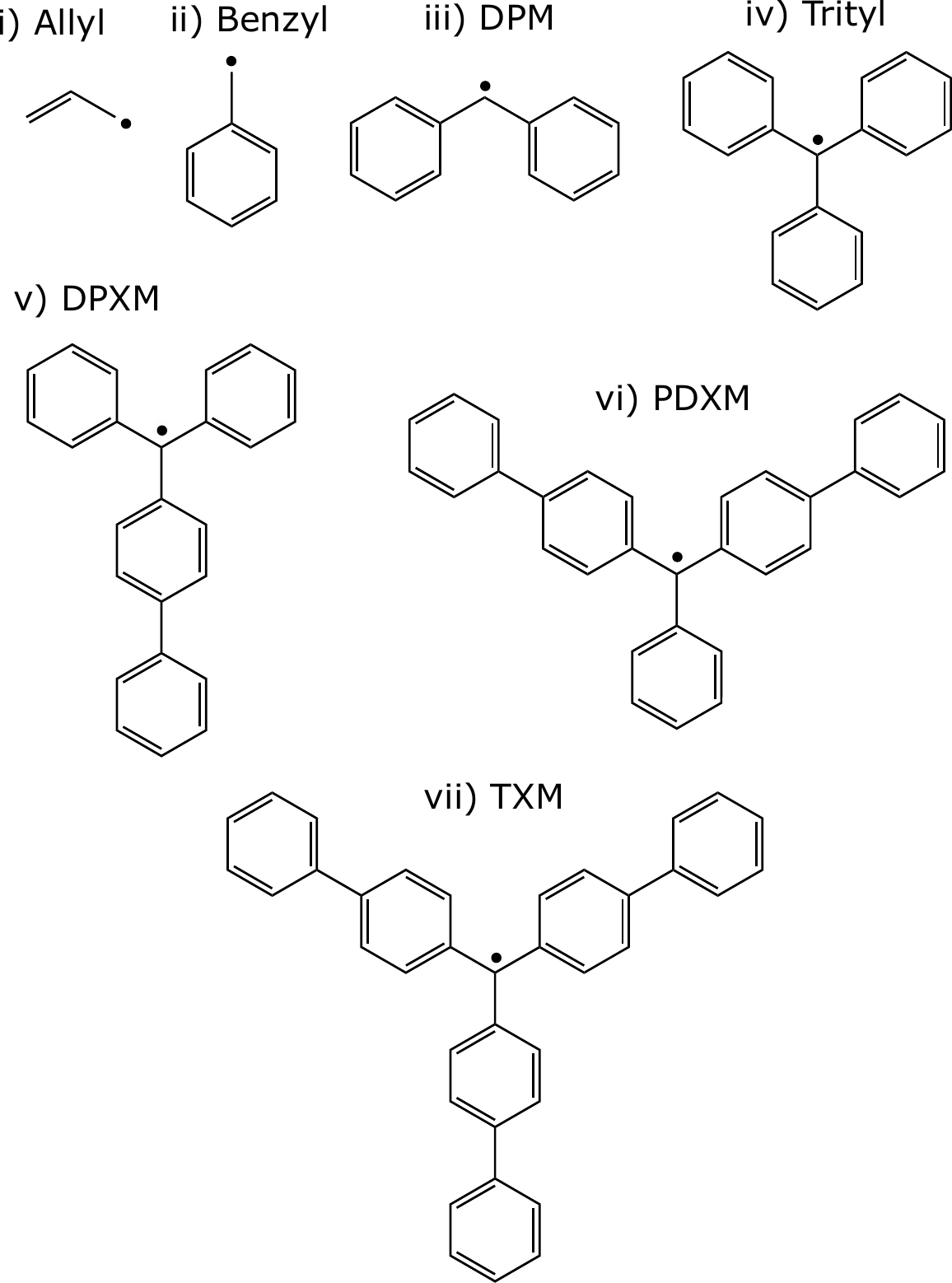}
\caption{The odd alternant hydrocarbons considered in this paper. The abbreviations were used for the following radicals: DPM -- diphenylmethyl, DPXM -- diphenyl-p-xenylmethyl, PDXM -- phenyl-di-p-xenylmethyl and TXM -- tri-p-xenylmethyl.}
\figl{mols}
\end{figure}

We now consider the accuracy, speed and spin purity of ExROPPP.

\subsubsection{Accuracy}
\paragraph{The allyl, benzyl and diphenylmethyl radicals.}
The ExROPPP and GMC-QDPT simulated spectra for the these radicals are presented in \figr{allyl} and \figr{benz_dpm} and their spectroscopic data are presented in Tables I -- VII in the supplementary information. The number of bands and their energies and relative intensities are consistent between the two theoretical methods and with the experimental data. 

\begin{figure}
\includegraphics[width=0.5\textwidth]{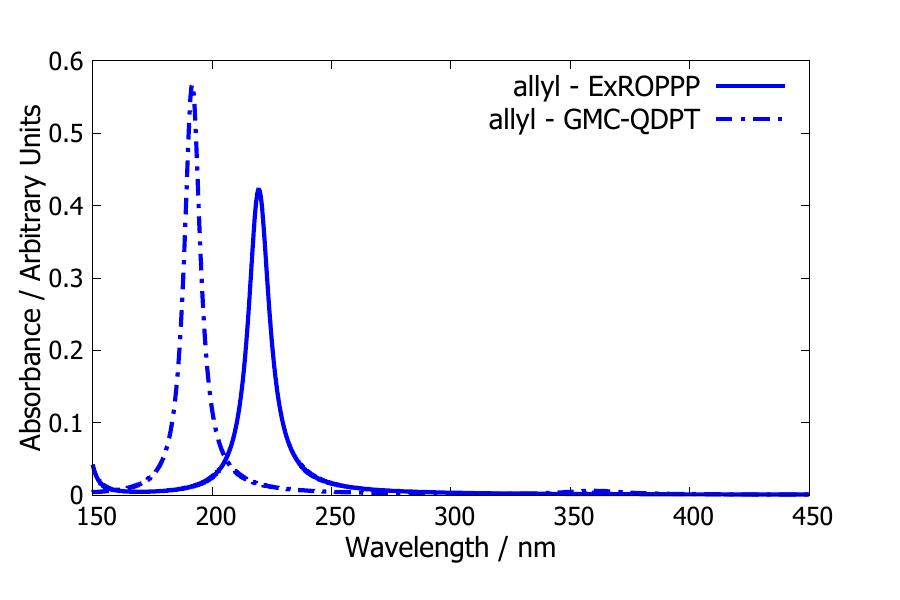}
\caption{UV-Visible vertical absorption spectra of the allyl radical simulated by ExROPPP (solid blue) and GMC-QDPT (dot-dashed blue).}
\figl{allyl}
\end{figure}

\begin{table}[h]
\caption{Excitation energies of the allyl radical calculated by ExROPPP and GMC-QDPT compared with experimental data.}
\begin{tabular}{c|c|c|c}
State & $E_{\textrm{ExROPPP}}$ (eV) & 
$E_{\textrm{GMC-QDPT}}$ (eV) & $E_{\textrm{exp.}}$ (eV)  \\ \hline
$1^2B_2$&3.05& 3.45&3.07 \\
$2^2B_2$&5.65 &6.47 & 5.00\\
\end{tabular}
\tabl{allyl}
\end{table}

For the allyl radical, in the $C_{2v}$ point group, the $1^2B_2$ state ($D_1$) is composed of $\ket{\Psi_{01}^-}$ in both GMC-QDPT and ExROPPP and appears at 3.07 eV (404 nm) in the experiment\cite{sch00a} and at 3.45 eV (360 nm) and 3.05 eV (407 nm) in GMC-QDPT and ExROPPP calculations respectively, showing a good agreement especially for ExROPPP. The experimental intensity of the $1^2B_2$ state is known to be very weak for other studies,\cite{ton00a,cur66a} which is in agreement with the low oscillator strengths computed by GMC-QDPT and ExROPPP of 0.005 and 0 respectively. The $2^2B_2$ state, $\ket{\Psi_{01}^+}$ in GMC-QDPT and ExROPPP, appears at 5.00 eV (248 nm) in the experiment\cite{sch00a} and 6.47 eV (192 nm) and 5.65 eV (220 nm) respectively in GMC-QDPT and ExROPPP, showing a close agreement. $2^2B_2$ is observed as a bright absorption in another study\cite{wat70a} which is in agreement with GMC-QDPT and ExROPPP with oscillator strengths of 0.567 and 0.422 respectively.

\begin{figure}
\includegraphics[width=0.5\textwidth]{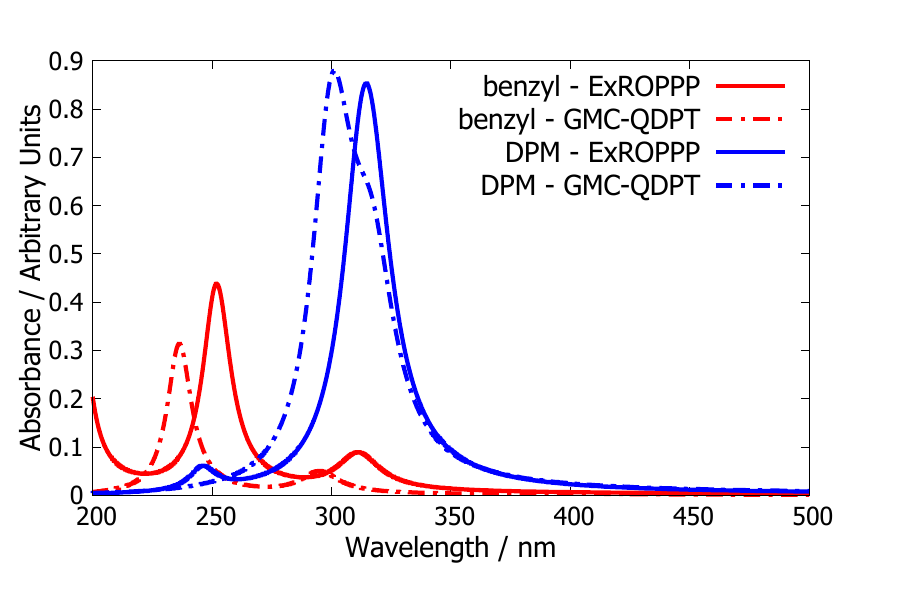}
\caption{UV-Visible vertical absorption spectra of benzyl (red) and diphenylmethyl (blue) radicals simulated by ExROPPP (solid lines) and GMC-QDPT (dot-dashed lines).}
\figl{benz_dpm}
\end{figure}

\begin{table}[h]
\caption{Excitation energies of the benzyl radical calculated by ExROPPP and GMC-QDPT compared with experimental data.\cite{joh68a}}
\begin{tabular}{c|c|c|c}
State & $E_{\textrm{ExROPPP}}$ (eV) & 
$E_{\textrm{GMC-QDPT}}$ (eV) & $E_{\textrm{exp.}}$ (eV)  \\ \hline
$1^2A_2$  &  2.89& 3.05 & 2.76  \\
$2^2A_2$  &3.98 & 4.20 &4.00\\
$4^2B_2$  & 4.92& 5.25& 4.77
\end{tabular}
\tabl{benzyl}
\end{table}

\begin{table}[h]
\caption{Excitation energies of the diphenylmethyl radical calculated by ExROPPP and GMC-QDPT compared with experimental data. }
\begin{tabular}{c|c|c|c}
State & $E_{\textrm{ExROPPP}}$ (eV) &  $E_{\textrm{GMC-QDPT}}$ (eV) & $E_{\textrm{exp.}}$ (eV)  \\ \hline
 $1^2A$&2.64 & 2.56&2.37 \\
 $4^2A$& 3.95&4.13 &3.69
\end{tabular}
 \tabl{dpm}
\end{table}

The weak $1^2A_2$ state ($D_1$) of the benzyl radical is seen around 2.76 eV (450 nm).\cite{joh68a} $1^2A_2$ is predicted to be at 3.05 eV (407 nm) in GMC-QDPT and 2.89 eV (430 nm) in ExROPPP and similar to the allyl radical is predicted to be completely dark by ExROPPP (composed of  $\ket{\Psi_{01}^-}$), with a very small oscillator strength predicted by GMC-QDPT. 
 The $3^2B_2$ state ($\ket{\Psi_{02}^-}$ in ExROPPP), which appears at 4.99 eV (248 nm) with $f\approx0$ in GMC-QDPT and 4.50 eV (275 nm) with $f=0$ in ExROPPP, is not seen in experiment likely due to it having a vanishingly small intensity. The two bright states of the benzyl radical, $2^2A_2$ and $4^2B_2$, which are composed of $\ket{\Psi_{01}^+}$ and $\ket{\Psi_{02}^+}$ in ExROPPP respectively are seen in the experimental spectra at 4.00 eV (310 nm) and 4.77 eV (260 nm) respectively.\cite{joh68a} These states are also well captured by GMC-QDPT ($2^2A_2$ at 4.20 eV (295 nm) and $4^2B_2$ at 5.25 eV (236 nm)) and ExROPPP ($2^2A_2$ at 3.98 eV (311 nm) and $4^2B_2$ at 4.92 eV (252 nm)) with particularly good agreement in ExROPPP. The relative intensities of these transitions are also in agreement between the two theoretical methods with the $4^2B_2$ state predicted to be brighter than the $2^2A_2$ by GMC-QDPT and ExROPPP, which is also consistent with the experimental spectrum.\cite{joh68a} While the agreement between ExROPPP and experiment is slightly better for the bright states than dark states, the accuracy of ExROPPP for the benzyl radical is overall as good as if not better than GMC-QDPT. 
 
 In the GMC-QDPT calculations, the orbitals 1 ($2b_2$) and 2 ($1a_2$) are swapped in energetic order compared to ExROPPP and the excited states seem to be composed of different excitations, however, on swapping the orbital indices 1 and 2 in the GMC-QDPT excitations, the states are the same as those in ExROPPP. To avoid confusion, we have omitted the compositions of the GMC-QDPT states here and these results can be found in the supplementary information.

The results for the benzyl radical are also supported by early theoretical results.\cite{bin55a,kru69a} In particular, Ref.~\citenum{bin55a} shows the same energy ordering and relative intensities of the excited states for the benzyl radical: $1^2A_2$ (weak absorption), $2^2B_2$ (weak absorption), $2^2A_2$ (strong absorption), $3^2B_2$ (strong absorption, corresponds to $4^2B_2$ in ExROPPP/GMC-QDPT). Ref.~\citenum{bin55a} only considered the first two orbitals above and below the SOMO in the benzyl radical, derived from the splitting of the $1^2E$ and $2^2E$ levels in benzene, so did not account for the extra dark state $3^2B_2$ in ExROPPP which involves orbitals 3 and 3$^\prime$ in the benzyl radical, so the energetic ordering in both methods is essentially the same for the states captured by both.

For DPM, which belongs to the $C_2$ point group, the $1^2A$ state ($D_1$) appears at 2.37 eV (523 nm) in the experiment\cite{bro85a}, and at 2.56 eV (485 nm) and 2.64 eV (470 nm) respectively in GMC-QDPT and ExROPPP and is primarily composed of $\ket{\Psi_{01}^-}$ in ExROPPP. This state has a weak absorption (370 $\textrm{dm}^3 \textrm{mol}^{-1} \textrm{cm}^{-1}$)\cite{bro85a} in the experiment and is dark in GMC-QDPT and ExROPPP. The bright $4^2A$ state is seen at 3.69 eV (336 nm) in experiment\cite{bro85a} and its energy is well reproduced by GMC-QDPT and ExROPPP with values of 4.13 eV (300 nm) and 3.95 eV (314 nm) respectively and is composed of $\ket{\Psi_{01}^+}$ in ExROPPP. It has a high measured intensity of 31000 $\textrm{dm}^3 \textrm{mol}^{-1} \textrm{cm}^{-1}$\cite{bro85a} which is in agreement with the large oscillator strengths calculated by GMC-QDPT and ExROPPP of 0.714 and 0.781 respectively. Similarly as for the benzyl radical, the ExROPPP calculated energy of the bright state is in closer agreement with experiment than for the dark ($D_1$) state. 
The energetic ordering of the doublet states in DPM is not the same in ExROPPP and GMC-QDPT due to near degeneracies and a high density of states at low energy. 

\paragraph{The triphenylmethyl radical and its p-xenyl- derivatives.}
\begin{figure}
\includegraphics[width=0.5\textwidth]{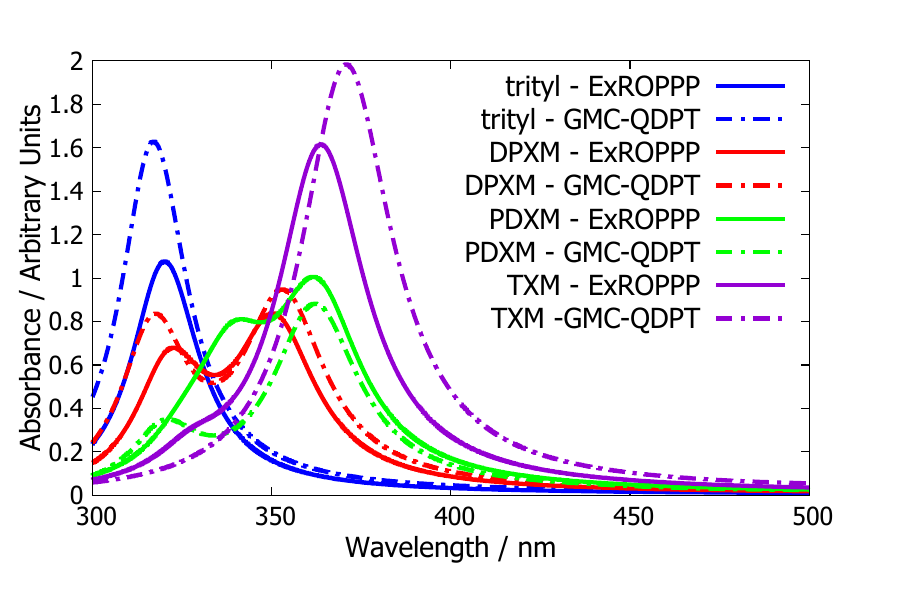}
\caption{UV-Visible vertical absorption spectra of the trityl radical and its derivatives DPXM, PDXM and TXM simulated by ExROPPP (solid lines) and GMC-QDPT (dot-dashed lines), showing good agreement between the two methods.}
\figl{tripxen}
\end{figure}
\begin{table}[h]
\caption{Excitation energies of the trityl radical calculated by ExROPPP and GMC-QDPT compared with experimental data. }
\begin{tabular}{c|c|c|c}
State & $E_{\textrm{ExROPPP}}$ (eV) &  $E_{\textrm{GMC-QDPT}}$ (eV) & $E_{\textrm{exp.}}$ (eV)  \\ \hline
 $1^2E$ &2.74 &2.63 & 2.41\\
 $2^2E$ &3.88 &3.91 &3.60
\end{tabular}
 \tabl{trityl}
\end{table}

\begin{table}[h]
\caption{Excitation energies of the DPXM radical calculated by ExROPPP and GMC-QDPT compared with experimental data. }
\begin{tabular}{c|c|c|c}
State & $E_{\textrm{ExROPPP}}$ (eV) &  $E_{\textrm{GMC-QDPT}}$ (eV) & $E_{\textrm{exp.}}$ (eV)  \\ \hline
 $2^2B$ &2.56  &2.40  &2.07\\
 $3^2B$ & 3.53& 3.51& 3.31 \\
 $2^2A$ & 3.86& 3.91& 3.60\\
\end{tabular}
\tabl{dpxm}
\end{table}

\begin{table}[h]
\caption{Excitation energies of the phenyl-di-p-xenylmethyl (PDXM) radical calculated by ExROPPP and GMC-QDPT compared with experimental data. }
\begin{tabular}{c|c|c|c}
State & $E_{\textrm{ExROPPP}}$ (eV) &  $E_{\textrm{GMC-QDPT}}$ (eV) & $E_{\textrm{exp.}}$ (eV)  \\ \hline
  $1^2A$ & 2.51&2.33&2.03\\
 $2^2A$ &3.42 & 3.42&3.03  \\
 $3^2B$ &3.66 & 3.87& 3.45\\
\end{tabular}
\tabl{pdxm}
\end{table}

\begin{table}[h]
\caption{Excitation energies of the tri-p-xenylmethyl (TXM) radical calculated by ExROPPP and GMC-QDPT compared with experimental data. }
\begin{tabular}{c|c|c|c}
State & $E_{\textrm{ExROPPP}}$ (eV) &  $E_{\textrm{GMC-QDPT}}$ (eV) & $E_{\textrm{exp.}}$ (eV)  \\ \hline
  $1^2E$ &2.52 &2.30& 2.03\\
 $2^2E$ &3.41& 3.34&2.95 
\end{tabular}
\tabl{txm}
\end{table}
The experimental UV/Visible absorption spectra of the trityl (triphenylmethyl) radical and its p-xenyl derivatives: di-phenyl-p-xenylmethyl (DPXM), phenyl-di-p-xenylmethyl (PDXM) and tri-p-xenylmethyl (TXM) contain one or two intense absorption bands depending on the molecule's point symmetry and are seen around 300--500 nm. They also have one or two very weak bands in the region of 500--700 nm in the visible.\cite{chu54a} Trityl and TXM, in the $D_3$ point group, both have a single intense near-UV band owing to the $2^2E$ state which is composed mainly of the $\ket{\Psi_{01}^+}$ and $\ket{\Psi_{02}^+}$ states (excitations between the SOMO and the doubly-degenerate HOMO/LUMO) as seen in GMC-QDPT and ExROPPP calculations. DPXM and PDXM, which both belong to the $C_2$ group, have two intense bands owing to the $\ket{\Psi_{01}^+}$ and $\ket{\Psi_{02}^+}$ states due to the removal of their degeneracy in descending from $D_3$ to $C_2$ symmetry, one of which having the irreducible representation $A$ and the other $B$ (the assignment of $A$ or $B$ to the excitations above differs between the two radicals). The ExROPPP and GMC-QDPT simulated spectra of these four radicals are presented in \figr{tripxen} and their spectroscopic data are presented in Tables~\ref{tab:trityl} -- \ref{tab:txm} (and in more detail in Tables IV -- VII in the SI). 

There is a good agreement between ExROPPP and the experimental data for these four radicals, which is better for the bright UV `$+$' states as opposed to the dark lowest energy excited `$-$' states in the visible which are completely dark in ExROPPP (due to the radicals all being alternant) and which only have a very small extinction coefficient in the experimental spectra. For the bright transitions, the ExROPPP values are within 0.28 eV (25 nm) of the experimental value for trityl, 0.26 eV (24 nm) for DPXM, 0.39 eV (48 nm) for PDXM and 0.46 eV (56 nm) for TXM. It is suggested by Ref.~\citenum{chu54a} that the lowest excited state of the trityl and TXM radicals is of $A_1$ symmetry from analysis of fluorescence polarisation data, however we believe this not to be that case as this lowest excited state of the trityl radical has been shown to be a doubly-degenerate $E$ state in a later publication\cite{mur57a} and in our results. 
The results of GMC-QDPT calculations for these radicals are also in good agreement with ExROPPP and the relative energy ordering of the doublet states is the same in both methods.

The excitation energies and relative intensities of the lowest excited states for the triphenylmethyl radical are also largely supported theoretical studies based loosely on Ref.~\citenum{lon55a} where the first excited state in their results is a $^2E$ state predicted to have a weak absorption at 2.41 eV (515 nm) and their fifth excited state is a $^2E$ state predicted to have an intense absorption at 3.84 eV (322 nm) corresponding to $2^2E$ in our results.\cite{mur57a} The number of states is different in both methods however, the energetic ordering of the states captured by both ExROPPP and the method of Ref.~\citenum{mur57a} is the same. The energies in Ref.~\citenum{mur57a} were shifted such that the energy of the lowest excited state matched experimental data, whereas our results are not manipulated in any way to match experiment.

\paragraph{Quartet states}
The agreement between the GMC-QDPT and ExROPPP calculated energies for the quartet states is not so good as for the doublet states. For the allyl radical, the quartet state energy is 6.07 eV  in GMC-QDPT and 4.40 eV in ExROPPP. Similarly, for the benzyl radical, GMC-QDPT predicts energies of 4.42 eV and 5.42 eV for the $1^4B_2$ and $1^4A_2$ states respectively, but the ExROPPP values are 3.62 eV and 4.80 eV respectively, again seeming to underestimate the quartet energy. ExROPPP also seems to largely underestimate the quartet energies for the trityl radical for the $1^4A_2$ and $1^4E$ states when compared to the GMC-QDPT values. In addition, the relative energetic ordering of doublet and quartet states is different in the GMC-QDPT and ExROPPP results. One possible reason for this discrepancy is that the set of PPP parameters used in this paper (and PPP theory in general) have mainly been applied to calculating the energies of singlet-singlet transitions in closed-shell molecules and in particular, simulating their spectra, in which only those transitions which are bright are of importance.\cite{mat57a} Therefore, by analogy, one would expect these parameters to be more accurate for calculating the energies of excited doublet states and less accurate for (dark) excited quartet states in radicals, which is what we observe from our results. 

Moreover it is known to be very difficult to experimentally measure quartet state energies as quartet states are formally dark. Also in the case of odd alternant hydrocarbon radicals, the first quartet state is usually higher in energy than the lowest excited doublet state so it is not normally observed through intersystem crossing from the doublet state in the same way that a triplet state is observed via intersystem crossing from the singlet state in a closed-shell molecule.\cite{sha21a} As a side note, this is a useful property of such radicals which makes them desirable for making OLEDs, but also means that the quartet state is rarely observed.\cite{lon55a,bay70a,fuk90a} The experimental quartet energies for the odd alternant hydrocarbon radicals tested, to the best of our knowledge, are absent from the literature so it is impossible to comment on the accuracy of ExROPPP for quartet states with any certainty.  
 
Finally, on analysing the CI expansion of the quartet states calculated by ExROPPP, we notice that the quartet states for the trityl, DPXM, PDXM and TXM radicals have significant contribution from higher excitations ($i>2$, $j^\prime>2^\prime$) and therefore it is expected that their computed energies will be dependent on inclusion of such higher excitations.

\paragraph{Overall accuracy}
 \figr{comp} shows a comparison of the experimental excitation energies and those computed by the two theoretical methods. The ExROPPP calculated energies are all close to the experimental values. As mentioned earlier, it is difficult to comment on the accuracy of ExROPPP for the calculation of quartet state energies. The root mean squared (RMS) errors of the ExROPPP and GMC-QDPT results $E_i^\textrm{calc.}$ compared to the experimental excitation energies $E_i^\textrm{exp.}$ have been calculated using the formula
\begin{align}
\epsilon_\textrm{RMS} = \sqrt{\frac{1}{N}\sum_{i=1}^{N}(E_i^\textrm{calc.}- E_i^\textrm{exp.})^2}.
\end{align}
The $\epsilon_\textrm{RMS}$ values for the ExROPPP and GMC-QDPT results are presented in \tabr{rms}. It can be seen that ExROPPP calculates excitation energies with a similar accuracy to GMC-QDPT for the doublet states of the alternant hydrocarbon radicals chosen. We must stress that the results of ExROPPP computations presented here are the raw results which are in no way shifted to agree with experiment (or GMC-QDPT) and the same parameters are used as in previous studies on closed-shell molecules.\cite{hel19a,gre22a} Computations using the bases of excitations and of spin-adapted states yield identical results. 

\begin{table}
\caption{RMS errors of i) ExROPPP and ii) GMC-QDPT calculated excited state energies compared to the experimental values and iii) ExROPPP calculated energies compared to GMC-QDPT for the odd alternant hydrocarbon radicals presented in this paper. Numbers without brackets are for all states, whereas numbers inside brackets are calculated for doublet states only.}
\tabl{rms}

\begin{tabular}{c|c}
& $\epsilon_\textrm{RMS}$ / eV \\ \hline
GMC-QDPT vs Exp. & (0.48)\\
ExROPPP vs Exp. & (0.33)\\
ExROPPP vs GMC-QDPT & 0.63 (0.50) 
\end{tabular}
\end{table}
\begin{figure}
\includegraphics[width=0.5\textwidth]{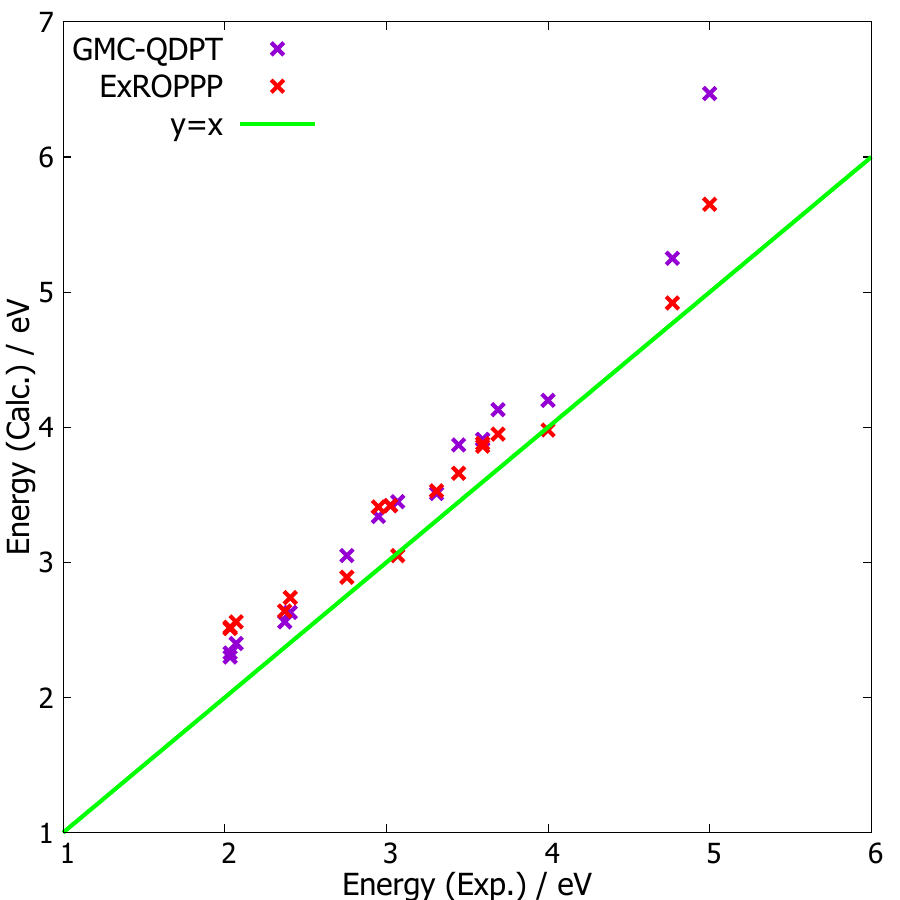}
\caption{Comparison of ExROPPP and GMC-QDPT excitation energies to experiment. No results have been shifted to agree with experiment and the ExROPPP results are produced using the same parameterization as has previously been used for closed shell molecules. ExROPPP is found to be as accurate as GMC-QDPT when compared to the experimental energies for doublet states.}
\figl{comp}
\end{figure} 

In general, the key features of the low lying electronic states of radicals are well captured by the simple and computationally inexpensive ExROPPP method. For the alternant hydrocarbon radicals studied, the lowest lying electronic state above the ground state is always a doublet $D_1$ and always has a very weak absorption in the experiment. This state is always $\ket{^2\Psi_{0i}^-}$ at the ExROPPP level of theory and always dark\cite{lon55a,bay70a}, but at the higher level of theory in GMC-QDPT it may have a very small oscillator strength.\cite{lon55a,bay70a} 
The doublet states $\ket{^2\Psi_{0i}^+}$ are always responsible for the low energy intense bands. There are higher lying doublet states of the type $\ket{^{2\mathrm{S}}\Psi_{i}^{j^\prime}}$ which are also bright, however for the molecules tested they lie far into the UV and are of less relevance to the optical properties. Furthermore, it is shown by ExROPPP calculations that only a small number excited states, at most $\ket{\Psi_{01}^+}$, $\ket{\Psi_{02}^+}$ and $\ket{\Psi_{03}^+}$, are responsible for the significant features of the UV/Visible spectrum for these molecules, involving transitions between the SOMO and the orbitals $1$, $2$, $3$, $1^\prime$, $2^\prime$ and $3^\prime$. 

\subsubsection{Spin-purity}
We calculated $\langle S^2\rangle_i$ and the degree spin-contamination $\Delta\langle S^2\rangle_i$ for each of the ExROPPP states $i$. We find that $\Delta\langle S^2\rangle_i=0$ for all ExROPPP states. However, in the basis of all single excitations only (ROPPP-CIS), $\Delta\langle S^2\rangle_i\neq0$ for all states (apart from for the allyl radical where only some states are contaminated, but this is a special case and not the trend), and $\Delta\langle S^2\rangle_i$ is generally significantly large, as shown in \tabr{benzyl_comp}. See Section V C in the SI for the calculation of spin-contamination.

\begin{table}[h]
\caption{Demonstration of spin-contamination $\Delta\langle{S^2}\rangle$ resulting from the use of a basis of all single-excitations. The excited states calculated by a) ROPPP-CIS (CIS with restricted open-shell PPP reference with all single excitations) and b) ExROPPP for the benzyl radical  ($C_{2v}$ point group) are presented and compared to the results of GMC-QDPT calculations and experimental data.\cite{joh68a} $1B_2$ is the ground state.}
\tabl{benzyl_comp}
\begin{tabular}{c|c|c|c|c|c|c|c|c}
State & $\langle{S^2}\rangle_\textrm{a)}$ & $\langle{S^2}\rangle_\textrm{b)}$ & $\Delta\langle{S^2}\rangle_\textrm{a)}$ & $\Delta\langle{S^2}\rangle_\textrm{b)}$ & $E_\textrm{a)}$ (eV) & $E_\textrm{b)}$ (eV) &  $E_{\textrm{GMC-QDPT}}$ (eV) & $E_{\textrm{exp.}}$ (eV)  \\ \hline
$1B_2$ & 0.789& 0.75&0.276 & 0& n/a&n/a &n/a & n/a\\
$1A_2$ & 1.016& 0.75& 0.679  &0 &2.99& 2.89& 3.05 &2.76\\
$2A_2$ &0.866 & 0.75&0.467 &0 &3.96 &3.98 & 4.20 & 4.00\\
$4B_2$ &0.761&0.75 &0.146 &0 & 4.83& 4.92& 5.25& 4.77
\end{tabular}
\end{table}

\subsubsection{Speed}
The speed of the ExROPPP method was compared to that of the accurate post-Hartree-Fock method, GMC-QDPT for the radicals studied. These results are shown in \tabr{sped}. 
GMC-QDPT calculations on all radicals were performed using GAMESS-US using 2 cores on a desktop computer and all ExROPPP calculations were performed using Python 3 using the same desktop computer and utilized 8 cores. The timings from ExROPPP presented here are for proof of concept purposes only since the ExROPPP code has not been optimised and was coded in Python; however, it clearly offers a substantial reduction in computation time compared to GMC-QDPT. We leave optimisation of the ExROPPP code and rigorous benchmarking for future work. We defined the speed-up factor as $t_\textrm{GMC-QDPT}/t_\textrm{ExROPPP}$.
\begin{table}[h]
\caption{Comparison of the speed of the ExROPPP method with GMC-QDPT. GMC-QDPT calculations are for doublet calculations only, and are at least two orders of magnitude slower than ExROPPP. Timings given are total CPU time.}
\tabl{sped}
\begin{tabular}{c|c|c|c}
Molecule&$t_\textrm{GMC-QDPT}$ / s &$t_\textrm{ExROPPP}$ / s & Speed-up factor\\ \hline
Allyl& 9.26&	0.0231&402\\
Benzyl& 1560&0.157	&	9970\\
DPM& 22000&	0.511	&43000\\
Trityl& 8670&	1.58&5500	\\
DPXM& 31800&	4.10&	7740\\
PDXM& 80800	&8.74	&9250\\
TXM& 163000 &	18.1&	9020
\end{tabular}
\end{table}

\clearpage

\subsection{Alternacy rules}
\label{sec:alt_rules}
In this section we apply the algebra of XCIS in the basis of spin-adapted \emph{plus} and \emph{minus} states to alternant hydrocarbon radicals. Using this method we can deduce simple and widely applicable rules for the interaction of the excited states of alternant hydrocarbon radicals which may guide spectral interpretation and molecular design of emissive radicals.

 It has long been known that due to pseudoparity, the excited states of closed-shell alternant hydrocarbons can be factored into \emph{plus} and \emph{minus} combinations, and that a further set of Hamiltonian interaction and dipole moment selection rules apply to them in addition to the usual rules concerning symmetry and spin.\cite{par56a} These rules were originally derived by Pariser considering the ground state and single-excitations and using the approximations of NDO and zero resonance integrals between non-nearest-neighbouring atoms. While the idea of pseudoparity has been known for a long time to also apply to radicals,\cite{lon55a} and some of Pariser's rules have been explored in special cases, to the best of our knowledge there has yet to be an attempt to derive a general set of rules for alternant hydrocarbon radicals. In this section we test Pariser's rules to see if they also apply to radicals. We apply the equations for the XCIS method for radicals mentioned above to derive the Hamiltonian and dipole moment interaction elements of \emph{plus} and \emph{minus} states, as it is the anolog for radicals of CIS in closed-shell systems.

In order to get a complete picture of the interaction of the excited states of radicals we need to consider \emph{plus} and \emph{minus} combinations of all XCIS states. The \emph{plus} and \emph{minus} states with one open-shell have already been worked out for us by Longuet-Higgins and Pople and are given in Section~\ref{sec:bkgnd} \eqr{ps0}.\cite{lon55a} However, as far as we are aware, Longuet-Higgins and Pople did not consider pseudoparity adapted states of three open-shells and we must now define them. The doublet and quartet states with three open-shells also come in degenerate pairs, and we can make the following linear combinations of these pairs of states
\bse
\begin{align}
\ket{^{2\mathrm{S}}\Psi_{ij}^\pm} &= \frac{1}{\sqrt 2}(\ket{^{2\mathrm{S}}\Psi_i^{j^\prime}} \pm \ket{^{2\mathrm{S}}\Psi_j^{i^\prime}}), \eql{ps2s} \\
\ket{^{2\mathrm{T}}\Psi_{ij}^\pm} &= \frac{1}{\sqrt 2}(\ket{^{2\mathrm{T}}\Psi_i^{j^\prime}} \mp \ket{^{2\mathrm{T}}\Psi_j^{i^\prime}}), \eql{ps2t} \\
\ket{^4\Psi_{ij}^\pm} &= \frac{1}{\sqrt 2}(\ket{^4\Psi_i^{j^\prime}} \pm \ket{^4\Psi_j^{i^\prime}}) \eql{ps4}
\end{align}
\ese
where $i \neq j$. When $i=j$, \eqr{ps2s}--\eqr{ps4} are no longer valid and only one state $\ket{^X\Psi_i^{i^\prime}}$ exists (where $X=2S$, 2T or 4).
Note that we have defined the $\ket{^{2\mathrm{T}}\Psi_{ij}^\pm}$ linear combinations with the opposite signs to the other two pairs of states: i.e. for $\ket{^{2\mathrm{S}}\Psi_{ij}^\pm}$ and $\ket{^4\Psi_{ij}^\pm}$, the \emph{plus} state has a `$+$' sign and the \emph{minus} state has a `$-$' sign in their expansions in \eqr{ps2s} and \eqr{ps4} respectively. However, for $\ket{^{2\mathrm{T}}\Psi_{ij}^\pm}$, the \emph{plus} state has a `$-$' sign and \emph{minus} state a `$+$' sign. The reason for this seemingly inconsistent definition will become apparent when we make the approximations relating to alternant hydrocarbons.

Now that we have defined a complete basis of pseudoparity and spin adapted states, we can derive the exact matrix elements for the XCIS Hamiltonian $\mathrm{\mathbf{H_{\textrm{XCIS}}^p}}$ in this basis, using the equations in Tables VIII and IX in SI Section II, discussed above. These are given in SI Section II C in Table X. We also derive the exact dipole moments $\mathrm{\mathbf{M_{\textrm{XCIS}}^p}}$ in the basis of \emph{plus} and \emph{minus} states which are presented in Table XIV in SI Section II G.

We then consider the effect of the same approximations as Pariser did on the XCIS Hamiltonian and dipole moments, in addition to the assumption that the Fock matrix is diagonalized, i.e. $F_{i,j} = \delta_{ij}\epsilon_i$ (in other words, this requires that the orbitals used in forming the Hamiltonian are the converged molecular orbitals). After making this series of approximations, the matrix elements simplify greatly and many of which vanish. The XCIS Hamiltonian $\mathrm{\mathbf{H_{\textrm{XCIS}}^a}}$ and dipole moment operator $\mathrm{\mathbf{M_{\textrm{XCIS}}^a}}$ in the basis of pseudoparity-and-spin-adapted states under the aforementioned approximations are given in Tables XI and XV respectively in Sections II D and II H of the SI. From this algebra clear rules for the excited state spectra of alternant radical hydrocarbons emerge which are similar but not identical to those for closed-shell molecules.

Firstly, we find that there is zero Hamiltonian interaction between \emph{plus} and \emph{minus} states of alternant hydrocarbon radicals. This has been previously shown to be true for closed-shell alternant hydrocarbons.\cite{par56a} In the case of radicals though there is the aforementioned caveat that we have to define the $\ket{^{2\mathrm{T}}\Psi_{ij}^\pm}$ states with the opposite signs in their expansion in terms of $\ket{^{2\mathrm{T}}\Psi_i^{j^\prime}}$ and $\ket{^{2\mathrm{T}}\Psi_j^{i^\prime}}$, and additionally we notice that $\ket{^{2\mathrm{T}}\Psi_{i}^{i^\prime}}$ states behave as \emph{minus} states whereas $\ket{^{2\mathrm{S}}\Psi_{i}^{i^\prime}}$ and $\ket{^{4}\Psi_{i}^{i^\prime}}$ behave as \emph{plus} states. We find that for radicals, the ground state behaves like a \emph{minus} state and only interacts with other \emph{minus} states. In closed-shell alternant hydrocarbons this is also true, however, due to Brillouin's theorem only higher than singly-excited \emph{minus} states may interact with the ground state. On the other hand, a weaker version of Brillouin's theorem applies for radicals where singly-excited states of the type $\ket{^2\Psi_{0i}^-}$ may mix with the ground state but all other purely singly-excited states are forbidden from interacting with the ground state ($\ket{^{2\mathrm T}\Psi_i^{i^\prime}}$ and $\ket{^{2\mathrm T}\Psi_{ij}^-}$ interact with $\ket{^2\Psi_0}$ but have doubly-excited character). The Hamiltonian interaction rules for radicals are illustrated pictorially below in \figr{pmmixing} and numerically in Table XI in SI Section II D.

\begin{figure}
\centering
\caption{Interaction of the electronic excited states of an alternant hydrocarbon radical via the Hamiltonian $\hat H$. Lines drawn between states represent an interaction via $\hat H$. States may interact if and only if they are in the same quadrant.}
\includegraphics[width=.5\textwidth]{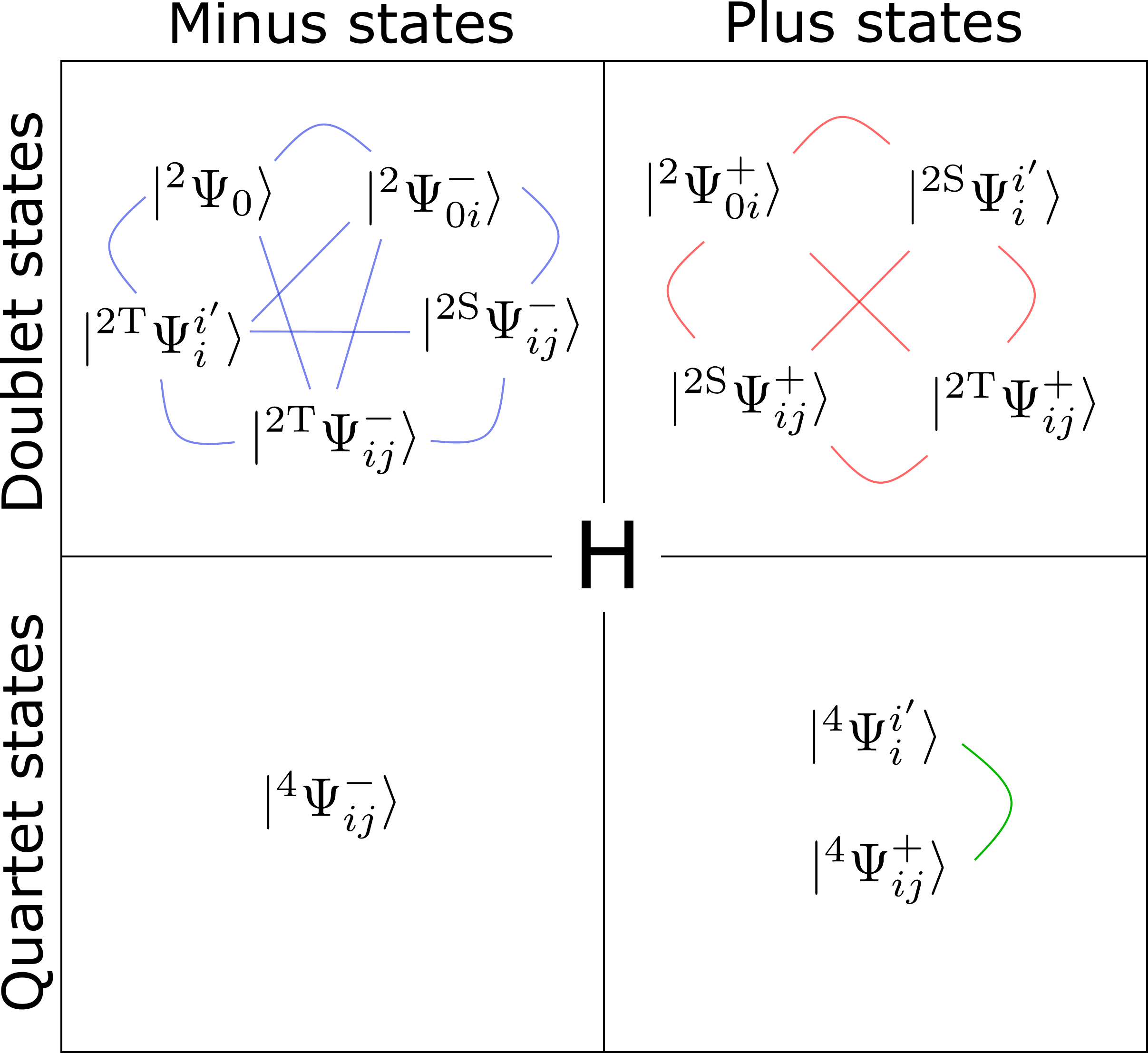}
\figl{pmmixing}
\end{figure}

Furthermore, Pariser showed that for closed-shell molecules, the Hamiltonian in the basis of singlet \emph{minus} states $\ket{^1\Psi_{ij}^-}$ is exactly the same as the Hamiltonian in the basis of triplet \emph{minus} states $\ket{^3\Psi_{ij}^-}$. While this is an interesting observation, we find no corresponding rule for radicals: no two non-zero Hamiltonian matrix elements are in general the same.

Secondly, we find that for radicals the dipole moments between two \emph{plus} states or between two \emph{minus} states are exactly zero. This selection rule applies on top of the more general spin selection rule and any selection rules imposed by point group symmetry.\cite{atk11a,sza89a} This pseudoparity selection rule has been previously known for closed-shell alternant hydrocarbons.\cite{par56a}
Consequently the transition dipole moment between two states may be non-zero if and only if the states are of opposite pseudoparity and equal multiplicity and regardless of whether the alternant hydrocarbon is closed-shell or a radical. It follows that the ground-state and any XCIS states of an alternant hydrocarbon have zero permanent dipole moment. The pseudoparity selection rule for radicals is illustrated pictorially below in \figr{dip_mixing} and numerically in Table XV in SI Section II H.

All of the above rules hold even after configuration interaction. That is, there exist four sets of states grouped by pseduoparity (\emph{plus} or \emph{minus}) and multiplicity (singlet/doublet or triplet/quartet) where any member of one set does not interact via the Hamiltonian with any member of any other set, before and after configuration interaction. Furthermore, states of the same pseudoparity and/or opposite multiplicity have zero dipole moment between one another before and after configuration interaction. 

Specifically concerning their dipole moment from the ground state $\ket{\Psi_0}$ (which is a \emph{minus} state of the lowest multiplicity), \emph{plus} states of the same multiplicity may be bright and have a non-zero transition dipole moment where as \emph{minus} states are always dark, even after configuration interaction. There are however cases where \emph{plus} states of the same multiplicity as the ground state may be completely dark (such as $\ket{^{2\mathrm{T}}\Psi_{ij}^+}$) or may have vanishingly small transition dipole moments, however, this is not in violation of any of the above rules. As a result of the dipole pseudoparity selection rule for alternant hydrocarbons and it holding after configuration interaction, it is usually (but \emph{not} always) the case that the lowest energy electronic excited state of an alternant hydrocarbon radical is completely dark at this level of theory\cite{mor61a,mur57a} and expected to have a vanishingly small dipole moment in experiment and in calculations at higher levels of theory.\cite{wat70a,joh68a,mcc60a,bro85a,arn92a,chu54a,abd20a} The reason for this is that the lowest state is usually a doublet \emph{minus} state of the form $\ket{\Psi_{i0}^-}$.\cite{lon55a,hud21a} The lowest excited state cannot be $\ket{\Psi_{i0}^+}$ as there will always be a corresponding \emph{minus} state of lower energy.\cite{lon55a,hud21a} If the first excited state is not $\ket{\Psi_{i0}^-}$ then it must contain at least some contribution from configurations of three open-shells. Inspecting the relevant matrix elements in SI Table XI, we find that $\ket{^4\Psi_{ij}^+}$ is always the lowest energy configuration out of all of those with open-shells in $i$, 0 and $j^\prime$ and $\ket{^4\Psi_{i}^{i^\prime}}$ is always the lowest energy configuration out of all of those with open-shells in $i$, 0 and $i^\prime$, so it is likely but not certain that in the case that $\ket{\Psi_{i0}^-}$ is not lowest in energy, the lowest energy state will be a quartet state which will also be dark. 

\begin{figure}
\centering
\caption{Interaction of the electronic excited states of an alternant hydrocarbon radical via the dipole moment operator $\hat\mu$. Lines drawn between states represent an interaction via $\hat \mu$. Only states with the same multiplicity \emph{and} opposite pseudoparity may interact through the dipole moment operator.}
\includegraphics[width=0.5\textwidth]{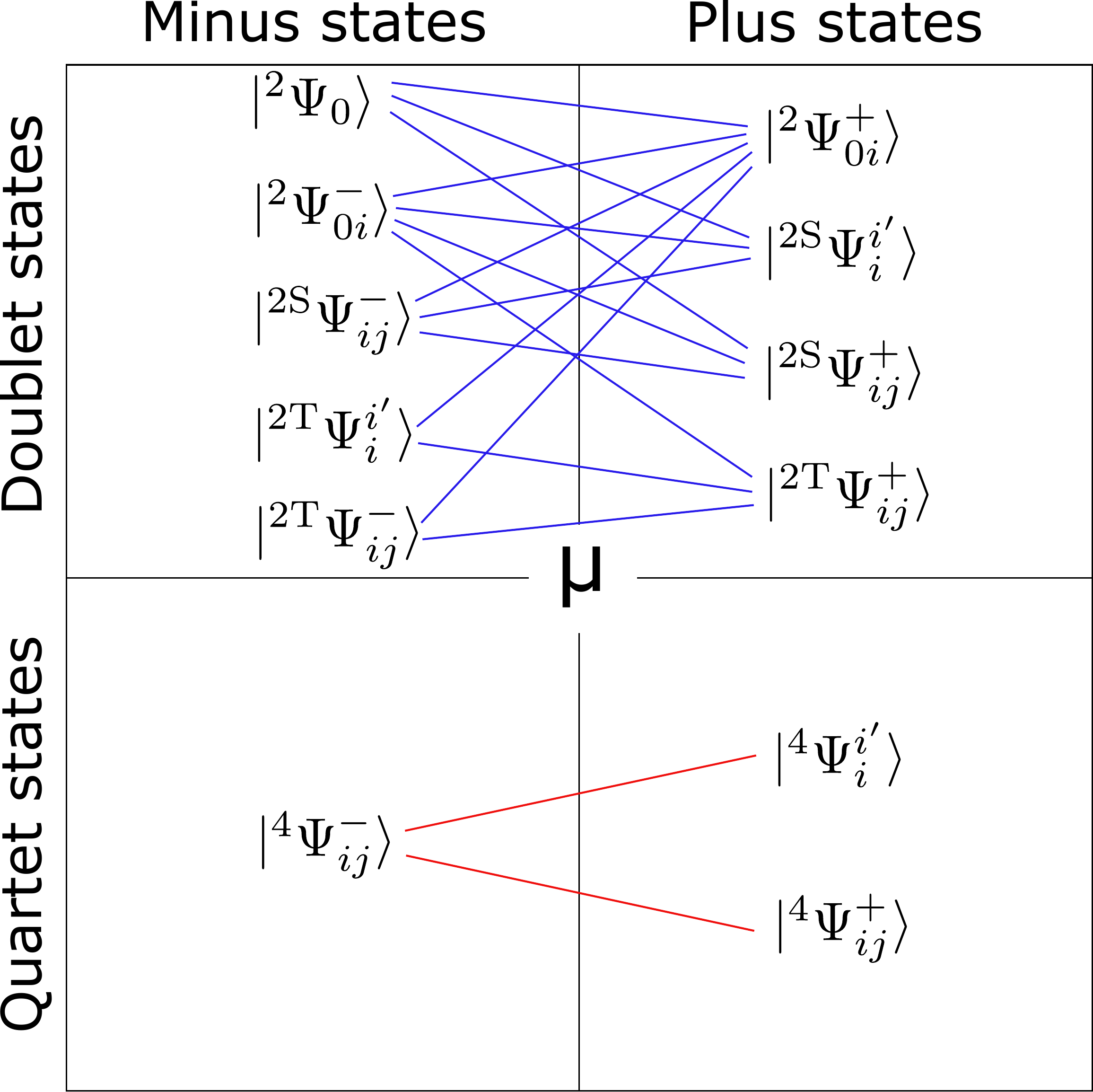}
\figl{dip_mixing}
\end{figure}

For practical purposes, the alternacy rules presented here mean that the UV-Vis spectrum of an alternant hydrocarbon radical will show a very weak absorption at low energies (corresponding to $D_1$ that will usually be $\ket{^2\Psi_{10}^-}$) and an intense absorption at higher energies, (usually corresponding to $\ket{^2\Psi_{10}^+})$. Since emission is usually from the lowest excited state by Kasha's rule, emission from alternant radicals is likely to be slow and outcompeted by nonradiative decay. Consequently, organic radicals for OLEDs should not be alternant hydrocarbons. These rules are consistent with, but stronger than, previous design rules that emissive radicals should be non-alternant\cite{abd20a,hel21a}. It was previously known that $D_1$ would usually be $\ket{^2\Psi_{10}^-}$ and dark, but this did not exclude the possibility of $\ket{^2\Psi_{10}^-}$ mixing with, and borrowing intensity\cite{rob67a} from, higher-lying bright excitations. However, the alternacy rules derived here show that $\ket{^2\Psi_{10}^-}$ can mix only with other minus excitations which also have zero dipole moment with the ground state, such that after configuration interaction $\ket{^2\Psi_{10}^-}$ will remain dark.

In addition, although the alternacy rules derived here strictly speaking only hold for purely hydrocarbon radicals, they are likely to hold qualitatively for chlorinated radicals such as the tris-(2,4,6-trichlorophenyl)-methyl (TTM) and perchlorotriphenylmethyl (PTM) radicals where the high electronegativity of chlorine compared to carbon and the mismatch between its 3p orbitals and the 2p orbitals of carbon mean that it is unlikely to participate significantly in the electronic structure in the conjugated system.\cite{abd20a,hud21a} 

Comparing the above alternacy rules with the numerical results from ExROPPP calculations on alternant hydrocarbons presented in Section~\ref{sec:exroppp_results} we see that the results of ExROPPP are entirely consistent with these rules. This can be seen in Tables I--VII in Section I of the SI, where the energies, oscillator strengths and CI expansion coefficients of the ExROPPP states are tabulated.
\clearpage
\begin{table}[h]
\caption{Novel excited state interaction rules derived for alternant hydrocarbon radicals compared with those already existing for closed-shell alternant hydrocarbons.}
\tabl{alt_rules}
\begin{tabular}{c|c|c}
Rule&Radical &  Closed-shell \\ \hline
1 & All states belong to one of two & All states belong to one of two \\
& pseudoparity groups, \emph{plus} or \emph{minus}. & pseudoparity groups, \emph{plus} or \emph{minus}.\\
2 & Degeneracy: & Degeneracy:\\
& $E(^2 \Psi_{\bar i}^{\bar0}) = E(^2\Psi_0^{i^\prime})$&$E(^1\Psi_i^{j^\prime}) = E(^1\Psi_j^{i^\prime})$\\
& $E(^{2\mathrm{S}}\Psi_i^{j^\prime}) = E(^{2\mathrm{S}}\Psi_j^{i^\prime})$& $E(^3\Psi_i^{j^\prime}) = E(^3\Psi_j^{i^\prime})$ \\ 
 & $ E(^{2\mathrm{T}}\Psi_i^{j^\prime}) =  E(^{2\mathrm{T}}\Psi_j^{i^\prime})$ &\\
 & $ E(^{4}\Psi_i^{j^\prime}) =  E(^{4}\Psi_j^{i^\prime})$ &\\
  3 & \emph{Plus} and \emph{minus} states: & \emph{Plus} and \emph{minus} states: \\
 &$\ket{^2\Psi_{0i}^\pm} = \frac{1}{\sqrt 2}(\ket{^2\Psi_{\bar{i}}^{\bar{0}}} \pm \ket{^2\Psi_0^{i^\prime}})$&$\ket{^1\Psi_i^{j^\prime,\pm}} = \frac{1}{\sqrt2}(\ket{^1\Psi_i^{j^\prime}} \pm \ket{^1\Psi_j^{i^\prime}})$  \\
& $\ket{^{2\mathrm{S}}\Psi_{ij}^\pm} = \frac{1}{\sqrt 2}(\ket{^{2\mathrm{S}}\Psi_i^{j^\prime}} \pm \ket{^{2\mathrm{S}}\Psi_j^{i^\prime}})$& $\ket{^3\Psi_i^{j^\prime,\pm}} = \frac{1}{\sqrt2}(\ket{^3\Psi_i^{j^\prime}} \pm \ket{^3\Psi_j^{i^\prime}})$\\
 &$\ket{^{2\mathrm{T}}\Psi_{ij}^\pm} = \frac{1}{\sqrt 2}(\ket{^{2\mathrm{T}}\Psi_i^{j^\prime}} \mp \ket{^{2\mathrm{T}}\Psi_j^{i^\prime}})$&$\ket{^1\Psi_i^{i^\prime}}$ and $\ket{^3\Psi_i^{i^\prime}}$ are \emph{plus} states. \\ 
 & $\ket{^4\Psi_{ij}^\pm} = \frac{1}{\sqrt 2}(\ket{^4\Psi_i^{j^\prime}} \pm \ket{^4\Psi_j^{i^\prime}})$&   \\
  & $\ket{^{2\mathrm{S}}\Psi_i^{i^\prime}}$ and $\ket{^4\Psi_i^{i^\prime}}$ are \emph{plus} states but & \\  
  & $\ket{^{2\mathrm{T}}\Psi_i^{i^\prime}}$ is a \emph{minus} state. &\\
 4  & $\ket{^2\Psi_0}$ is a \emph{minus} state. & $\ket{^1\Psi_0}$ is a \emph{minus} state.\\
5 &There is zero Hamiltonian interaction & There is zero Hamiltonian interaction \\
& between \emph{plus} and \emph{minus} states. &between \emph{plus} and \emph{minus} states.\\
6  & N/A &$\bra{^1\Psi_{ij}^-}\hat H\ket{^1\Psi_{kl}^-} = \bra{^3\Psi_{ij}^-}\hat H\ket{^3\Psi_{kl}^-} $ \\
7 &For any two arbitrary states $\ket{\Psi_u}$ and $\ket{\Psi_v}$, & For any two arbitrary states $\ket{\Psi_u}$ and $\ket{\Psi_v}$, \\& $\bra{\Psi_u}\hat \mu\ket{\Psi_v}$ may be non-zero if and only if& $\bra{\Psi_u}\hat \mu\ket{\Psi_v}$ may be non-zero if and only if\\
&  $\ket{\Psi_u}$ and $\ket{\Psi_v}$ have equal multiplicity and& $\ket{\Psi_u}$ and $\ket{\Psi_v}$ have equal multiplicity and\\ 
& opposite pseudoparity.& opposite pseudoparity.\\
8  & $D_1$ state usually $\ket{^2\Psi_{01}^-}$ which is always &$S_1$ state usually $\ket{^1\Psi_1^{1^\prime}}$ which is usually\\
&dark. & bright. \\
9 & Rules 1, 5, and 7 hold after configuration & Rules 1, 5, and 7 hold after configuration\\
 & interaction.& interaction.
\end{tabular}
\end{table}
\clearpage

\section{Conclusions}
In this article we have combined previous ideas that adding certain double excitations to a radical CIS calculation can lead to spin purity\cite{lon55a,mau96a,hel21a} with Pariser-Parr-Pople theory \cite{par53a,pop53a,pop55a,par56a, kou67a} to obtain a computational method for the low-lying electronically excited states of hydrocarbon radicals. This method, which we call ExROPPP, reproduces experimental doublet excitation energies and intensities for a series of odd alternant hydrocarbon radicals accurately and with minimal computational effort, with a similar degree of accuracy to General Multiconfigurational Quasi-Degenerate Perturbation Theory (GMC-QDPT), an accurate multiconfigurational SCF method. In addition, using the approximations of PPP theory we have derived widely-applicable rules for alternant hydrocarbons which should aid in the prediction, assignment and interpretation of their spectra. 

There are many avenues for future research. The proof-of-concept calculations in this paper are for hydrocarbon radicals (for which there already exists tested parameterization) and we plan to extend this method to radicals containing nitrogen and chlorine such as the breakthrough fluorescent radical TTM-3NCz\cite{ai18a}, sulfur containing radicals such as those based on thiophene\cite{mur23a} and non-alternant and ionic radicals. There are possibilities of using machine learning\cite{dra21a} to optimize ExROPPP parameterization, especially for heteroatoms and for quartet states, given that our initial results suggest that ExROPPP is less accurate for quartet state energies than doublet state energies. The ExROPPP method could be extended to radicals with more than one unpaired electron in their ground state\cite{stu19a}, and the alternacy rules we have derived for monoradicals may be extendable to these systems. It is our hope that ExROPPP can be extended to accurately and quickly simulate a wide range of conjugated organic radicals, becoming a useful tool for rapid spectral determination. This can then be used to design highly efficient and emissive radicals\cite{hel21a} for applications such as OLEDs\cite{ai18a} and quantum computing.\cite{gor23a}

\section*{Supplementary Material}
The supplementary material consists of tables containing the detailed energies, oscillator strengths/intensities and composition of excited states calculated by ExROPPP and GMC-QDPT and obtained from experimental data. Also deposited here are molecular geometries 
 used for the calculations, and any other details of the computational methods used in this paper.

\section*{Acknowledgements}
The authors thank Emrys W.\ Evans for helpful suggestions on the research. TJHH acknowledges a Royal Society University Research Fellowship URF\textbackslash R1\textbackslash 201502 and a startup grant from University College London.

\section*{Data Availability Statement}
The data that supports the findings of this study are available within the article and its supplementary material.

\section*{AIP Publishing Credit Line}
This article may be downloaded for personal use only. Any other use requires prior permission of the author and AIP Publishing. This article appeared in J. Chem. Phys. 160, 164110 (2024) and may be found at https://doi.org/10.1063/5.0191373

\bibliography{roppp_final_1_short}
\end{document}

%% file: abbrev.tex
\newcommand{\bra}[1]{\langle #1 |}
\newcommand{\ket}[1]{| #1 \rangle}

\newcommand{\bmr}{\mathbf{r}}

\newcommand{\eqr}[1]{Eq.~\eqref{eq:#1}}

\newcommand{\eql}[1]{\label{eq:#1}}
\newcommand{\figr}[1]{Fig.~\ref{fig:#1}}
\newcommand{\figl}[1]{\label{fig:#1}}

